\documentclass[aps,reprint,preprintnumbers,twocolumn,floatfix,superscriptaddress,pra,showpacs,longbibliography]{revtex4-1}
\usepackage{physics}
\usepackage{natbib}
\usepackage{amsmath,amssymb}
\usepackage{graphicx}
\usepackage[dvipsnames]{xcolor}
\usepackage[caption=false,subrefformat=parens]{subfig}
\usepackage{float}
\usepackage[pdftex]{hyperref}

\begin{document}
	\title{Quantum-Assisted Telescope Arrays}
	\author{E. T. Khabiboulline}
	\email{ekhabiboulline@g.harvard.edu}
	\affiliation{Department of Physics, Harvard University, Cambridge, Massachusetts 02138, USA}
	\author{J. Borregaard}
	\affiliation{Department of Physics, Harvard University, Cambridge, Massachusetts 02138, USA}
	\affiliation{QMATH, Department of Mathematical Sciences, University of Copenhagen, 2100 Copenhagen \O, Denmark}
	\author{K. \surname{De Greve}}
	\author{M. D. Lukin}
	\affiliation{Department of Physics, Harvard University, Cambridge, Massachusetts 02138, USA}
	
	\begin{abstract}
	Quantum networks provide a platform for astronomical interferometers capable of imaging faint stellar objects. In a recent work [arXiv:1809.01659], we presented a protocol that circumvents transmission losses with efficient use of quantum resources and modest quantum memories. Here we analyze a number of extensions to that scheme. We show that it can be operated as a truly broadband interferometer and generalized to multiple sites in the array. We also analyze how imaging based on the quantum Fourier transform provides improved signal-to-noise ratio compared to classical processing. Finally, we discuss physical realizations including photon-detection-based quantum state transfer.  
	\end{abstract}
	
	\maketitle

	\section{Introduction}
	
	Telescope arrays boost the angular resolution in astronomical imaging~\citep{Lawson2000,Wootten2009}. By interfering the light collected at sites across the array, a synthetic aperture with resolution proportional to the length of the array and frequency of the light is realized~\cite{Thompson1999}. In practice, however, transmission losses limit the separation between sites. The resolution is then restricted, in particular if the light sources under study are weak, which is typically the case for imaging in the optical domain~\citep{brummelaar2005}. \citet{Tsang2011} demonstrated theoretically that in this setting a nonlocal measurement, such as in direct detection, is necessary for good performance. Alternatively, quantum entanglement can connect sites; quantum teleportation-based interference of the stellar light via distributed entangled photon pairs was initially proposed by \citet{Gottesman2012}. However, estimates of the necessary rate of entanglement distribution for such an approach suggested a high rate exceeding 100 GHz, which currently is not feasible. The introduction of quantum memories into the network offers a significant relaxation of this requirement. In Ref.~\cite{prl} we showed that the quantum state of the collected light can be compressed and stored nonlocally across the network, yielding an exponential reduction in the consumption of entangled resources, as compared to memoryless schemes. The necessary rate of entanglement distribution is reduced by several orders of magnitude, which opens up realistic prospects for employing near-term quantum networks~\citep{Simon2017,Humphreys2018} for high-resolution imaging in the optical domain.
	
\begin{figure}
		\centering
		\includegraphics[width=0.5\textwidth]{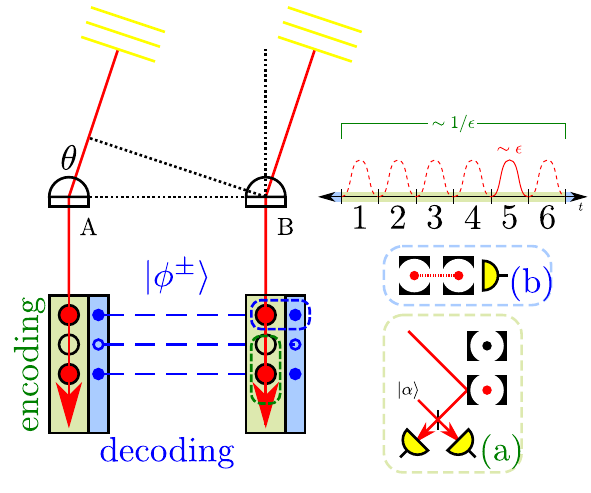}
		\caption{Memory-based interferometry scheme from Ref.~\cite{prl}. The quantum state of an incoming photon and associated information is stored nonlocally between telescope sites in a binary qubit encoding that can be decoded using preshared entangled pairs. Encoding operations are performed in time bins set by the detector bandwidth, followed by decoding after one photon is expected to have arrived. Physical realization may involve qubits housed in cavities: (a) Reflecting the photon off cavities, interfering with a coherent state, and detecting photons performs the encoding, while (b) decoding is done with qubit-qubit interactions followed by measurement.
			\label{fig:scheme}}
	\end{figure}

	In this article we further develop the scheme presented in Ref.~\cite{prl} (see Fig.~\ref{fig:scheme}) and analyze a number of possible extensions relevant to the setting of astronomical interferometry. In particular, we analyze how the limited bandwidth of typical quantum memories can be overcome by means of frequency splitting followed by efficient encoding. Next we describe how the original two-node scheme can be extended to a multiple-site array. We also study how processing the network's stored quantum data with a quantum Fourier transform improves imaging. Finally, Ref.~\cite{prl} suggests initially transferring the incoming optical modes to an auxiliary atomic qubit in a Raman-absorption scheme. Here we show how to eliminate the auxiliary atom, by reflecting the photons off cavities and interfering them with ancillary photonic states in a beam splitter, followed by photon detection.
	
	The paper is organized as follows. Section~\ref{sec:broadband} considers broadband light and coding of the frequency data. In Sec.~\ref{sec:multiple-telescope}, operation is generalized to arrays with more than two telescope sites, and the advantage of a quantum Fourier transform performed over the network is discussed in Sec.~\ref{sec:qft}. State transfer to memory based on photon detection is elaborated in Sec.~\ref{sec:state transfer}, with further considerations given in Appendixes~\ref{sec:pPort} and~\ref{sec:imperfectDetection}.
		
	\section{Broadband operation} \label{sec:broadband}
 	The essence of the scheme in Ref.~\cite{prl} is to transfer the quantum state of the photon to a logical qubit with a binary encoding of the arrival time. Digitization of time is set by the characteristic scale of the inverse detector bandwidth; this time bin contains an average number of $\epsilon$ photons (see Fig.~\ref{fig:scheme}). After encoding over $\sim$$1/\epsilon$ time bins such that one photon is expected, entanglement-assisted parity checks between the telescope sites determine its arrival time in a nondestructive manner. Crucially, this measurement projects out the vacuum component of the light and fixes the nonlocal quantum state. The phase information relevant for interferometry can then be processed without suffering from the vacuum noise that impairs local detection schemes~\cite{Tsang2011}. The binary code means that only $\log_2(M+1)$ memory qubits are needed per site, where $M\sim1/\epsilon$ is the number of time bins integrated over. Consequently, also only $\sim$$\log_2 1/\epsilon$ entangled pairs are consumed for the parity checks, giving a significant reduction in entanglement distribution rate compared to memoryless schemes.  
	
	Real stellar light has a broad frequency distribution. Meanwhile, interferometers typically have a small bandwidth~\cite{brummelaar2013} to avoid washing out spatial correlations. Furthermore, detectors operate over a narrow frequency band in order to ensure high-resolution imaging by, for example, avoiding dark counts~\cite{Hadfield2009}. The amount of light collected in the interferometer is then limited. Moreover, the broadband information of stellar light is potentially useful for astronomy~\citep{Stee2017}. From the point of view of near-term quantum networks, the multiplexing scheme developed here can compensate for slow gate time.
	
\begin{figure}
		\centering
		\includegraphics[width=0.5\textwidth]{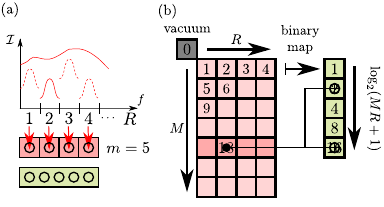}
		\caption{Efficient encoding of photon frequency and time information in $\log_2(MR+1)$ memory qubits. (a) The incoming photon arrives in one of $M$ time bins (see Fig.~\ref{fig:scheme}) and one of $R$ frequency bands. The excitation is transferred to a frequency-matched receiving qubit (red) and later stored in memory qubits (green). (b) The time-frequency data have $MR+1$ possibilities (+1 for vacuum), which can be mapped to $\log_2(MR+1)$ codewords in binary. An example of the encoding {\scriptsize CNOT} gates is shown for the fifth time bin and second frequency band. The number of detectors scales linearly in $R$, whereas the coding operation consumes resources logarithmic in both $M$ and $R$.
			\label{fig:frequencyStorage}}
	\end{figure}

	A generalization of the protocol to broadband operation is shown in Fig.~\ref{fig:frequencyStorage}. The incoming light is split into $R$ frequency bands. Quantum frequency conversion~\citep{kumar,Kambs2016,Dreau2018} enables operation at some convenient frequency, so that the photons can be stored in receiving atoms in a Raman-absorption scheme~\citep{cirac1997,boozer2007} (see Fig.~\ref{fig:frequencyStorage}(a)). We assume the incoming light to be weak ($\epsilon\ll1$) such that at most one photon arrives. Thus, similar to Ref.~\cite{prl}, $M\gg1$ time bins are integrated over in order to record, on average, one photon in any of the $R$ frequency bands. We wish to store both the time and frequency data of the photon in order to ensure interferometric operation. The same basic idea of binary encoding can be applied. Write the time bin $m$ and frequency band $r$ as one string $i=(m,r)$, which as a whole can be expressed in a binary expansion. If there are $M$ time bins and $R$ frequency bands, $\log_2(MR+1)$ codewords are needed (the +1 term accounts for the possibility of no photon arriving) (see Fig.~\ref{fig:frequencyStorage}(b)).
	
	Consider the encoding operation for the concrete case of a photon arriving in the fifth time bin and two possible frequency bands. From the discussion above, the code requires $\lceil\log_2(5\cdot 2+1)\rceil=4$ memory qubits. Assuming that the photon is equally likely to fall within either of the bands, we approximate the photonic state on two telescope sites as
	\begin{eqnarray}
	\rho&\approx&(1-\epsilon)\rho^{(0)}+\frac{\epsilon}{2}\rho_1^{(1)}+\frac{\epsilon}{2}\rho_2^{(1)}\,, \label{eq:multiband1} \\
	\rho^{(0)}&=&\ket{0,0}_1\bra{0,0}\otimes\ket{0,0}_2\bra{0,0}\,, \\
	\rho_1^{(1)}&=&\frac{1}{2}(\ket{0,1}_1\bra{0,1}+\ket{1,0}_1\bra{1,0}+g_1\ket{1,0}_1\bra{0,1} \nonumber \\
	&&+g_1^*\ket{0,1}_1\bra{1,0})\otimes\ket{0,0}_2\bra{0,0}\,, \\
	\rho_2^{(1)}&=&\ket{0,0}_1\bra{0,0}\otimes\frac{1}{2}(\ket{0,1}_2\bra{0,1}+\ket{1,0}_2\bra{1,0} \nonumber \\
	&&+g_2\ket{1,0}_2\bra{0,1}+g_2^*\ket{0,1}_2\bra{1,0})\,,
	\end{eqnarray}	
	where $\ket{1,0}_1$ denotes the photon arriving at the first telescope site in the first frequency band while $\ket{1,0}_2$ corresponds to the first site and second band. Thus, $\rho^{(0)}$ denotes vacuum in all modes and $\rho^{(1)}$ are single-photon states. Since the incoming light is thermal~\cite{Tsang2011}, we have assumed that there are no correlations between the frequency bands. Spatial correlations result in the coherences $g_1$ for the first frequency band and $g_2$ for the second. As described in Ref.~\cite{prl}, the goal of the interferometer is to extract these coherences. The photonic state in Eq.~(\ref{eq:multiband1}) can be transferred to an atomic equivalent by a Raman-absorption scheme at each telescope site. Subsequent application of logical controlled-NOT ({\small CNOT} or {\small CX}) gates between the receiving atom and the four memory atoms at each site, followed by measurement of the receiving atom in the $X$ basis, establishes the transfer of the photon into memory. For our particular example, the logical {\small CNOT} corresponding to two frequency bands and the fifth time bin makes the transformation
	\begin{eqnarray}
	\ket{0,0}\ket{0,0}\ket{0000,0000}&\to&\ket{0,0}\ket{0,0}\ket{0000,0000}\,, \\
	\ket{1,0}\ket{0,0}\ket{0000,0000}&\to&\ket{1,0}\ket{0,0}\ket{1010,0000}\,, \\
	\ket{0,1}\ket{0,0}\ket{0000,0000}&\to&\ket{0,1}\ket{0,0}\ket{0000,1010}\,, \\
	\ket{0,0}\ket{1,0}\ket{0000,0000}&\to&\ket{0,0}\ket{1,0}\ket{1011,0000}\,, \\
	\ket{0,0}\ket{0,1}\ket{0000,0000}&\to&\ket{0,0}\ket{0,1}\ket{0000,1011}\,,
	\end{eqnarray}
	where the notation is
	\begin{eqnarray}
	&&\overbrace{\ket{0,0}}^{\text{freq.~1}}\overbrace{\ket{1,0}}^{\text{freq.~2}}\overbrace{\ket{1011,0000}}^{\text{memories}}\,, \\
	&&\overbrace{\lvert\underbrace{101}_{\text{time}}\underbrace{1}_{\text{freq.}},0000\rangle}^{\text{memories}}\,,
	\end{eqnarray}
	i.e., the first four qubits denote the state of the two receiving atoms at the respective sites separated into the two frequency bands while the remaining qubits are the memories. As before, we use commas to delimit qubits at separate sites. Within the memory, the first three qubits encode the time bin ($5\to101$), while the fourth encodes the frequency band. Time and frequency are encoded separately for simplicity, which does not incur a real penalty here.

\begin{figure}
		\centering
		\includegraphics[width=0.5\textwidth]{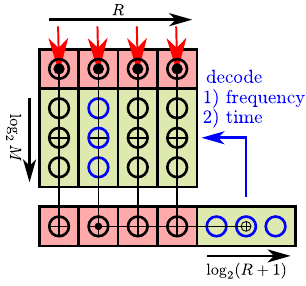}
		\caption{Encoding of the time bin ($m=5$) for each frequency band in parallel. To keep track of frequency, the receiving atom is copied to another atom before being reinitialized. Before decoding, these ancillary atoms are compressed to $\log_2(R+1)$ memory atoms (the nontrivial {\scriptsize CNOT} for frequency band $r=2$ is shown). Once the frequency is determined via nonlocal parity checks, only the corresponding time memory is decoded, so total entanglement expenditure scales logarithmically in both $M$ and $R$. Memory qubits scale as $R\log_2M$, to leading order.
			\label{fig:parallelFrequencyStorage}}
	\end{figure}

	Note that the logical {\small CNOT} gate described above requires each of the $R$ receiving qubits to interact with the same memory qubits, i.e., the encoding will happen sequentially. The operation needs to be fast compared to the detection bandwidth such that dead time is negligible. Instead, the encoding can be done in parallel by allotting each of the $R$ receiving qubits its own memory, which now encodes exclusively the time bin, as in Ref.~\cite{prl} (see Fig.~\ref{fig:parallelFrequencyStorage}). Another $R$ ancillary qubits are used to store the frequency information since the receiving qubits are reinitialized for each time bin. After $M$ time bins, the frequency information is first read out using $\log_2(R+1)$ entangled pairs: Compress the information stored in the $R$ ancillary qubits into $\log_2(R+1)$ qubits and then read them out through nonlocal parity checks as before. The arrival time is subsequently read out from the identified frequency band's memory using $\log_2M$ entangled pairs. This variant of our scheme has parallel operation over frequencies, at the expense of memories scaling as $R\log_2M$. Note, however, that the entanglement consumption still scales only logarithmically in $R$ and $M$.
	
	\section{Multiple-site array} \label{sec:multiple-telescope}
	So far, we have focused on two-site interferometry, but realistic astronomical interferometers require many nodes to reconstruct the stellar brightness distribution~\cite{brummelaar2013}. An array of sites with different spatial separation $x$ provides samples of the visibility $g(x)$. From these samples, the intensity distribution $\mathcal{I}(y)$ is obtained through a Fourier transform as specified by the van Cittert-Zernike theorem~\cite{Zernike1938}. Here we describe in detail how the scheme in Ref.~\cite{prl} can be generalized to networks with multiple nodes. We restrict the discussion to a single frequency band, but the extension to broadband operation is straightforward.

	For a weak source, we can model the light impinging on $N$ telescope sites with a density matrix~\cite{Tsang2011} $\rho\approx{(1-\epsilon)\rho^{(0)}}+\epsilon\rho^{(1)}$, where $\epsilon\ll1$, $\rho^{(0)}$ denotes vacuum in all modes, and
	\begin{equation*}
	\rho^{(1)}=\frac{1}{N}\left(
	\begin{array}{ccccc}
	1 & g_{0,1} & g_{0,2} & \ldots & g_{0,N-1}\\
	g_{1,0} & 1 & g_{1,2}  & \ldots & g_{1,N-1} \\
	g_{2,0} & g_{2,1} & 1 & \ldots & g_{2,N-1} \\
	\vdots & \vdots  &\vdots& \ddots & \vdots \\
	g_{N-1,0} & g_{N-1,1} & g_{N-1,2} &\ldots & 1
	\end{array}\right)\,,
	\end{equation*} 
	in the basis $\left\{\ket{1_i}\right\}$ where $i=\{0,\ldots,N-1\}$ and $\ket{1_i}$ is the state with a photon at the $i$th site and vacuum in the remaining modes. We have assumed that the probability for each telescope site to record the photon is the same and have defined $g_{i,j}=g(x_i-x_j)$ as the visibility at separation $x_i-x_j$, where $x_i$ is the position of the $i$th site; for simplicity of notation assume a linear array, distributed between $x=0$ and $x=d$, where $d$ is the maximal length. Note that $g_{i,j}=g_{j,i}^*$ and $g_{i,i}=1$. 
	
\begin{figure}
		\centering
		\includegraphics[width=0.5\textwidth]{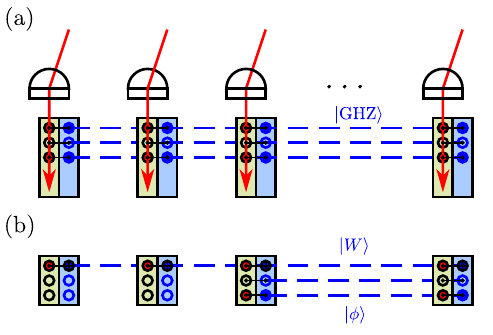}
		\caption{Interferometry in a multiple-site array. (a) The photon arrival data are decoded using $\log_2(M+1)$ GHZ states. A {\scriptsize CZ} gate is performed between the memory register at each site and the qubit of the GHZ state. For those registers corresponding to a photon, the phase of the GHZ state is flipped from $+$ to $-$, which can be read by measuring the qubits of the GHZ state in the $X$ basis. (b) For pairwise readout, acting on a shared $W$ state with local {\scriptsize CNOT} gates and measuring its qubits in the $Z$ basis projects the network state onto two sites, with probability $1-1/N$. Otherwise, the $W$ state is transformed into $\ket{0,\ldots,0}$ and measurement can be retried with another $W$ state. Two-side readout proceeds as before, using Bell states for the other registers. After pairwise visibilities are collected, a classical Fourier transform is applied to acquire the desired intensity distribution.
			\label{fig:network_ghz-w}}
	\end{figure}

	The photon is encoded at each site in the same way as for two-site operation, using the protocol of Ref.~\cite{prl}. After $M$ time bins of encoding, the arrival time of the single photon is again decoded with nonlocal parity checks. Only the particular registers flipped for the codeword corresponding to the photon's time bin will have odd parity. The Bell states $\ket{\phi^{\pm}}=(\ket{0,0}\pm\ket{1,1})/\sqrt{2}$ from Ref.~\cite{prl}, providing the nonlocality for the parity checks, are promoted to Greenberger-Horne-Zeilinger (GHZ) states of the form $\ket{\text{GHZ}^\pm}=(\ket{0,\ldots,0}\pm\ket{1,\ldots,1})/\sqrt{2}$ distributed across the array (see Fig.~\ref{fig:network_ghz-w}(a))~\footnote{Note that teleporting a single qubit around the array and performing local controlled-$Z$ gates would act similarly to the GHZ approach}. The qubits from parallel registers will have either even parity corresponding to the state $\ket{0,\ldots,0}$ or odd parity corresponding to a $W$ state $\ket{W}=\frac{1}{\sqrt{N}}\sum_i\ket{1_{i}}$. Performing local controlled-$Z$ ({\small CZ}) gates between the memory qubits in a register and the qubits of the GHZ state gives the transformation
	\begin{eqnarray}
	\ket{0,\ldots,0}\ket{\text{GHZ}^+}&\xrightarrow{N \times\text{CZ}}&\ket{0,\ldots,0}\ket{\text{GHZ}^+}\,,\\
	\ket{W}\ket{\text{GHZ}^+}&\xrightarrow{N \times\text{CZ}}& \ket{W}\ket{\text{GHZ}^-}\,.
	\end{eqnarray}
	Subsequently measuring the qubits of the GHZ state in the $X$ basis reveals the parity of the register without leaking information about photon location. Using $\log_2(M+1)$ GHZ states, one for each register, the arrival time can thus be decoded while preserving full interferometric operation. When a photon is recorded, all but one of the odd-parity registers are redundant and can be measured out in the $X$ basis, similar to the procedure in Ref.~\cite{prl}. The even-parity registers are all in a product state of $\ket{0}$s and can be traced out. Thus, the single-photon component of the photonic density matrix $\rho^{(1)}$ is mapped to an atomic equivalent at the telescope sites.
	
	The visibilities $g_{i,j}$ can be extracted one at a time using $W$ states, similar to the approach in Ref.~\cite{Gottesman2012}. A $W$ state is first distributed across the network. Next, local {\small CNOT} operations are performed between each memory qubit in a register (control) and the corresponding qubit from $\ket{W}$ (target). All qubits in $\ket{W}$ are subsequently measured in the $Z$ basis, with two possible outcomes. First, all qubits may be found in state $\ket{0}$, in which case $\rho^{(1)}$ is left intact and the procedure should be retried with a new $W$ state. This outcome happens with probability $1/N$. Second, with probability $1-1/N$, two qubits are in state $\ket{1}$ with the rest in state $\ket{0}$. If qubits $i$ and $j$ are found in state $\ket{1}$, then $\rho^{(1)}$ is transformed into
	\begin{multline}
	\frac{1}{2}\big(\ket{0,1}_{ij}\bra{0,1}+\ket{1,0}_{ij}\bra{1,0}\\+g_{i,j}\ket{1,0}_{ij}\bra{0,1}+g^*_{i,j}\ket{0,1}_{ij}\bra{1,0}\big)\otimes\rho^{(N-2)}_{0}\,,
	\end{multline} 
	where $\ket{0,1}_{ij}$ is the state with the memory qubit at the $i$th ($j$th) telescope in state $\ket{0}$ ($\ket{1}$). Here $\rho^{(N-2)}_{0}$ denotes that all other memory qubits are in state $\ket{0}$. As in Ref.~\cite{prl}, the visibility $g_{i,j}$ of this two-site state can be extracted by means of one-qubit measurements.

	The $W$ state operation may be done directly, to perform the parity check instead of using a GHZ state (see Fig.~\ref{fig:network_ghz-w}(b)). If the register has no excitation, then the $W$ state remains unchanged; otherwise, it transforms into one of the two possibilities described above. After collapse of the network state to two sites, the other registers can be processed as in Ref.~\cite{prl}, using Bell states.
	
	Following either of the above procedures, the visibilities in the array are sampled randomly, similar to the protocol of Ref.~\cite{Gottesman2012}. Repeating the procedure gives a distribution of samples across all possible pairwise combinations in the array. Fourier transforming this classical data then yields the intensity distribution of the source.
	
	\section{Quantum Fourier Transform} \label{sec:qft}
	
\begin{figure}
		\centering
		\includegraphics[width=0.5\textwidth]{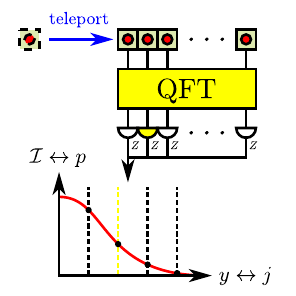}
		\caption{After transferring the memory qubits to one site via quantum teleportation, a quantum Fourier transform is applied. The measurement probabilities $p_j$ of finding the excitation at site $j$ map directly to the source intensity distribution $\mathcal{I}(y)$.
			\label{fig:network_qft}}
	\end{figure}

	The GHZ approach maintains coherence across the network, in the form of a nonlocal state $\rho^{(1)}$. The quantum data can be processed with a quantum Fourier transform (QFT), as was initially suggested in Ref.~\cite{Gottesman2012}. First transfer the $N$ qubits to one site via quantum teleportation, in order to perform all subsequent operations locally (see Fig.~\ref{fig:network_qft}); this step is more a matter of practicality than necessity. A QFT coherently interferes the off-diagonal elements of $\rho^{(1)}$, corresponding to the pairwise visibilities in the array, such that the intensity distribution $\mathcal{I}(y)$, where $y$ is the stellar coordinate, can be extracted from the resulting density matrix directly. Measurement noise only enters in the direct measurement of $\mathcal{I}(y)$, in contrast to the classical approach above where the visibilities are first sampled from the density matrix via measurement and then interfered in a classical Fourier transform (FT) to obtain $\mathcal{I}(y)$. The extra measurement noise will add in the FT, resulting in a more noisy estimate of $\mathcal{I}(y)$ than with the QFT.
	
	To quantify the possible gain of using a QFT, we assume a situation where the array sites are equally distributed along a line segment of length $d$. We can then label the $N-1$ sample points of the visibility as $g^{(j)}=g_{i,i-j}$, where $i>j$. Performing the QFT on the memories amounts to the operation
	\begin{equation}
	U_{\text{QFT}}\rho^{(1)}U^{\dagger}_{\text{QFT}}=\rho^{(\mathcal{I})}\,,
	\end{equation}	
	where the QFT unitary is
	\begin{equation} \label{eq:QFT}
	U_{\text{QFT}}(N)=\frac{1}{\sqrt{N}}\left(
	\begin{array}{ccccc}
	1 & 1 & 1 & \ldots & 1\\
	1 & \omega & \omega^2  & \ldots & \omega^{N-1} \\
	1 & \omega^2 & \omega^4  & \ldots & \omega^{2(N-1)} \\
	\vdots & \vdots  &\vdots& \ddots & \vdots \\
	1 & \omega^{N-1} & \omega^{2(N-1)}  & \ldots & \omega^{(N-1)^2}
	\end{array}\right)\,,
	\end{equation}	
	with $\omega=\text{exp}\left(2\pi i/N\right)$. Here $\rho^{(\mathcal{I})}$ has diagonal elements
	\begin{equation} \label{eq:rho_ft}
	\rho^{(\mathcal{I})}_{j,j}=\frac{1}{N}+\frac{2}{N^2}\sum_{k=1}^{N-1}(N-k)\Re\left\{g^{(k)}e^{2\pi i x_k y_j}\right\}\,,
	\end{equation}
	where $x_k=\frac{d}{N-1}k$ and $y_j=\frac{N-1}{Nd}j$. The diagonal elements thus directly correspond to estimates $\mathcal{I}^e(y_j)$ of $\mathcal{I}(y)$ for a finite number of sample points and where the visibilities $g^{(k)}$ have been weighted according to how much information is contained about them in $\rho^{(1)}$. For example, $g^{(N-1)}$ only appears once in $\rho^{(1)}$ while $g^{(1)}$ appears $N-1$ times. This structure is referred to as natural weighting in the literature and results in the minimum error in the estimate for $\mathcal{I}(y)$ for point sources~\cite{Thompson2007}. The measurement of $\mathcal{I}^e(y_j)$ simply consists of projecting onto the $Z$-basis states of $\rho^{(1)}$; the diagonal elements $\rho^{(\mathcal{I})}_{j,j}$ are precisely the probabilities $p_j$ of finding the excitation at site $j$. Assuming that $l\gg N$ samples of $\rho^{(1)}$ are measured, the variance of the QFT estimate will follow that of a multinomial distribution: $(\Delta\mathcal{I}_q^{e}(y_j))^2=\mathcal{I}^e(y_j)(1-\mathcal{I}^e(y_j))/l$. For the classical approach, where the same natural weighting as in Eq.~\eqref{eq:rho_ft} is employed in the FT, we can bound the variance of the estimate of the intensity distribution by $(\Delta\mathcal{I}_c^e(y_j))^2\geq1/l$ as a consequence of the standard propagation of errors~\cite{Ku1966} in the classical FT. 
	
	The advantage of the QFT in general depends on the number of sample points $N-1$ of the visibility and the actual intensity distribution being imaged. These factors determine the number of terms being coherently interfered in the QFT as opposed to incoherently interfered in a classical FT. For intuition, consider the example of a flat intensity distribution, corresponding to nonzero elements only on the diagonal of $\rho^{(1)}$. For an array size $N$, the density matrix contains $N$ diagonal elements, which are coherently summed in the QFT. The variance is proportional to $\mathcal{I}^e(y_j)$, which is normalized such that $\sum_i\mathcal{I}^e(y_i)=1$; for a flat distribution, $\mathcal{I}^e(y_j)=1/N$. With the classical approach, however, we have an incoherent sum and consequently, the variance is a factor of $N$ larger than for the QFT. Another illustrative example is the imaging of a point source. The spatial correlations are maximal and completely described by relative phases. After performing the QFT, a single qubit is flipped corresponding to the position of the point source, similar to how a lens, via coherent interference of the paths, focuses light from different directions onto different spots in the focal plane. Since the qubit is excited with unity probability, the variance is zero: The QFT perfectly identifies the position of the point source. In contrast, classical processing would result in fluctuations since the visibilities are measured and subject to shot noise.
	
	\section{Photon-detection-based state transfer} \label{sec:state transfer}
	
\begin{figure}
		\centering
		\includegraphics[width=0.5\textwidth]{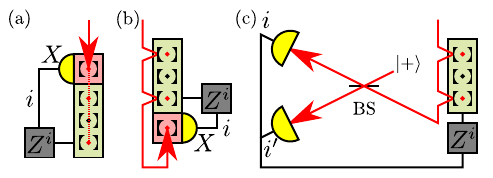}
		\caption{The stellar photon's state is transferred to memory in three ways. (a) The photon is absorbed by an auxiliary atom. Interactions with code-specified memory atoms realize {\scriptsize CNOT} gates. An $X$-basis measurement decouples the atom and imposes a phase correction on the memory to complete state transfer, as described in Ref.~\cite{prl}. (b) The {\scriptsize CNOT} gates are realized by reflecting the photon off cavities~\citep{Duan2004}. The photon is subsequently absorbed by an auxiliary atom to perform the $X$-basis measurement. (c) In photon-detection-based state transfer, measurement of the photonic qubit in the $X$ basis is approximately realized by interference at a beam splitter with an ancillary photonic state, followed by photon counting.
			\label{fig:stateTransferCombined}}
	\end{figure}

	In Ref.~\cite{prl} we suggested transferring the photonic excitation to an auxiliary atom through a Raman-absorption scheme. Atom-atom gates then realize the {\small CNOT} gates in the memory encoding (see Fig.~\ref{fig:stateTransferCombined}(a)). Note that the auxiliary atom can be reused for each time bin, with the requirement of fast two-qubit gates and measurement. Other methods of operation may be desirable depending on the experimental details. Instead of transferring the photonic excitation to an atom right away, the {\small CNOT} gates may be realized by reflecting the photon off cavities~\citep{Duan2004} specified by the binary code (see Fig.~\ref{fig:stateTransferCombined}(b)). As argued in Ref.~\cite{prl}, only the nontrivial {\small CNOT} operations that cause a bit flip should introduce error in order for the scheme to have efficient error accumulation. In this photon-atom gate implementation, the absence of a photon does not introduce any error on the atoms. The quantum state transfer to memory is completed by measuring the photonic qubit in the $X$ basis, where the measurement result determines a phase correction to be applied to the atomic state~\citep{prl}. Similar to before, the photon may be absorbed by an auxiliary atom, which is then straightforward to measure in the $X$ basis. However, quantum state transfer without the need for atomic absorption and measurement may be desirable: The experimental setup is simplified and fast photonic detection can be employed. We show that the photonic $X$-basis measurement can be approximately realized by local beam splitter interference.
	
	The purpose of the $X$-basis measurement is to erase the which-path information of the photon, allowing for a coherent transfer of the state to the atomic qubits up to a phase correction (similar to one-bit teleportation~\citep{Zhou2000}). The particular difficulty comes from the photonic qubit states corresponding to the absence $\ket{0}$ or presence $\ket{1}$ of a photon, for which a projection onto the $X$ basis, $\ket{\pm}=(\ket{0}\pm\ket{1})/\sqrt{2}$, is not readily available. In contrast, atomic qubits can be manipulated via spin rotations. An approximate rotated-basis measurement can nonetheless be achieved by mixing the incoming light with an ancillary photonic state on a 50:50 beam splitter and counting the photons at the outputs (see Fig.~\ref{fig:stateTransferCombined}(c)). For illustration purposes, we first describe how to emulate photonic $X$-basis measurements with an ancillary photonic superposition state $\ket{+}$. We then extend the results to the experimentally more feasible situation where the ancillary states are coherent states.
	
	The photonic state $\ket{+}$ has equal weight of vacuum and photon and can be used to obscure the absence or presence of the stellar photon when mixed with the incoming light in a beam splitter. If both photodetectors measure no photons or one measures two photons and the other zero (the Hong-Ou-Mandel effect), we can tell exactly whether a stellar photon arrived or not. However, if one detector measures no photons and the other detects one, then we cannot determine if the photon came from the locally injected light or the stellar source. There is a subtlety: Reflection off the beam splitter imparts an, in principle, distinguishable phase shift. Concretely, the beam splitter action is:
	\begin{eqnarray}
	\ket{0+}&\xrightarrow{\text{BS}}&\ket{0'_+}=\frac{1}{\sqrt{2}}[\ket{00}+\frac{1}{\sqrt{2}}(\ket{01}+\ket{10})]\,,\nonumber\\
	\ket{1+}&\xrightarrow{\text{BS}}&\ket{1'_+}=\frac{1}{\sqrt{2}}[\frac{1}{\sqrt{2}}(\ket{01}-\ket{10})\nonumber\\
	&&+\frac{1}{\sqrt{2}}(\ket{02}-\ket{20})]\,,
	\end{eqnarray}
	where, e.g., $\ket{0+}\equiv\ket{0}_1\otimes{\ket{+}}_2$ are the beam splitter modes. Including the memory qubits with logical states $\{\ket{\bar{0}},\ket{\bar{1}}\}$, the full state after the {\small CNOT} operation and beam splitter interference is $(\ket{0'_+}\ket{\bar{0}}+e^{i\theta}\ket{1'_+}\ket{\bar{1}})/\sqrt{2}$, where $\theta$ is the phase of the incoming light, assumed to be in state $(\ket{0}+e^{i\theta}\ket{1})/\sqrt{2}$. Postselecting on measuring only one photon in either of the detectors, we obtain the atomic state $(\ket{\bar{0}}-e^{i\theta}\ket{\bar{1}})/\sqrt{2}$ if the first detector clicks ($\ket{10}$ is measured) and $(\ket{\bar{0}}+e^{i\theta}\ket{\bar{1}})/\sqrt{2}$ if the second detector clicks ($\ket{01}$ is measured). Consequently, perfect state transfer is obtained if a logical phase-flip gate $Z$ is applied on the memory conditional on measuring a photon only in the first detector. The success probability of the operation is 1/2. 
	
	We propose mixing with the more practical coherent states $\ket{\alpha}=e^{-{\lvert\alpha\rvert}^2/2}\sum^\infty_{i=0} \frac{\alpha^i}{\sqrt{i!}}\ket{i}$, which can be readily produced classically. For a two-site telescope array, consider the transfer of a state of the form $\ket{\psi}=(\ket{0,1}+e^{i\theta}\ket{1,0})/\sqrt{2}$ to the atomic memories. After the logical {\small CNOT} operation with the memory atoms, the combined state will be $\ket{\Psi}=(\ket{0,1}\ket{\bar{0},\bar{1}}+e^{i\theta}\ket{1,0}\ket{\bar{1},\bar{0}})/\sqrt{2}$. Mixing the stellar light with a coherent state $\ket{\alpha}$ in a beam splitter at each site makes the transformation 
	\begin{eqnarray}
	\ket{0,1}\ket{\alpha,\alpha}&\xrightarrow{2\times\text{BS}}&\ket{0'_\alpha,1'_\alpha}\,, \\
	\ket{1,0}\ket{\alpha,\alpha}&\xrightarrow{2\times\text{BS}}&\ket{1'_\alpha,0'_\alpha}\,,
	\end{eqnarray}	
	where $\{\ket{0'_\alpha},\ket{1'_\alpha}\}$ each describes the two output modes of a beam splitter. Detection of $(i,i')$ photons at the first site and $(j,j')$ photons at the second site is specified by the measurement operator $\ket{ii',jj'}\bra{ii',jj'}$. The atomic state following the measurement is
	\begin{equation}
	\rho=\frac{1}{p(ii',jj')}\bra{\Psi}\ket{ii',jj'}\bra{ii',jj'}\ket{\Psi}\,,
	\end{equation}
	where $p(ii',jj')=\tr\{\bra{\Psi}\ket{ii',jj'}\bra{ii',jj'}\ket{\Psi}\}$ is the probability of measuring the combination $(i,i')$ photons at the first site and $(j,j')$ photons at the second site. The off-diagonal terms in $\rho$, describing the coherence between the two sites, are proportional to $(i-i')(j-j')$. Hence, a corrective $Z$ gate should be applied to the memory at each site based on which port detects more photons. 
	
\begin{figure}
		\centering
		\subfloat{\label{fig:tradeoff_probability1}\includegraphics[width=0.5\textwidth]{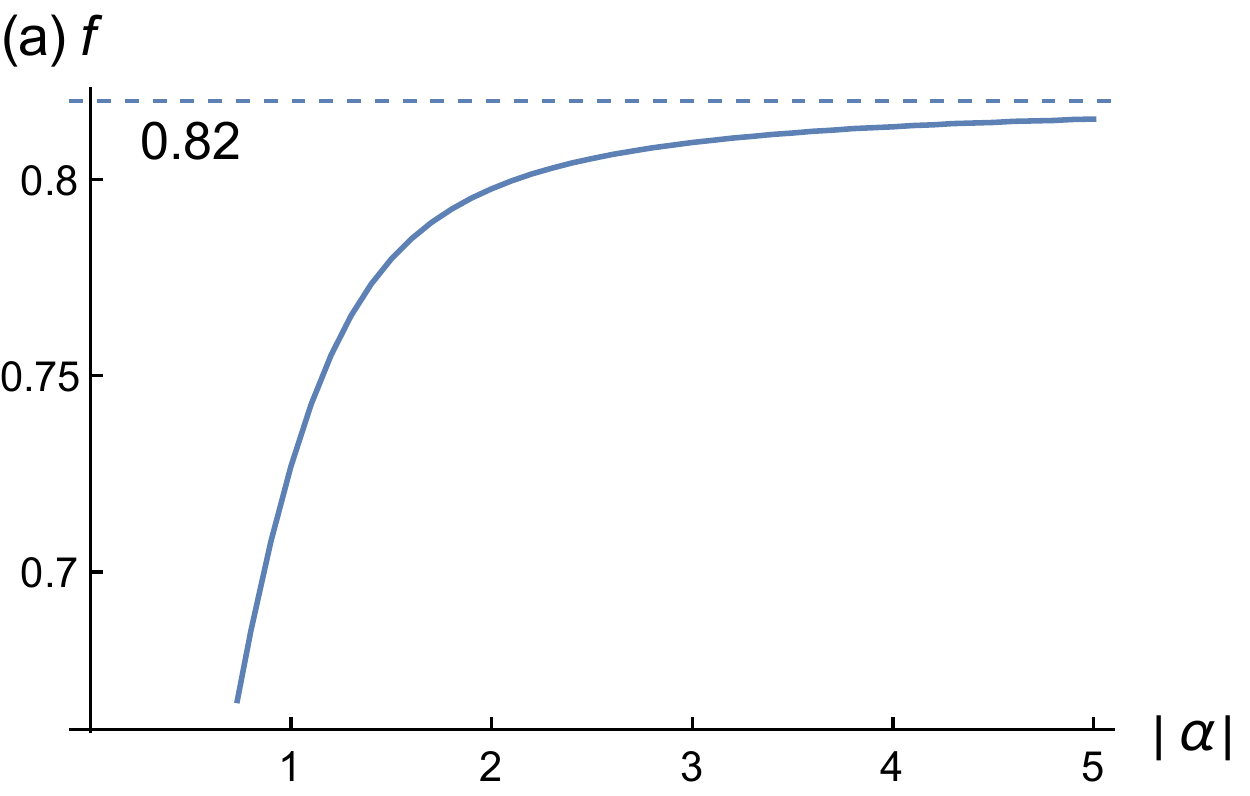}}
		\quad
		\subfloat{\label{fig:tradeoff_fidelity1}\includegraphics[width=0.5\textwidth]{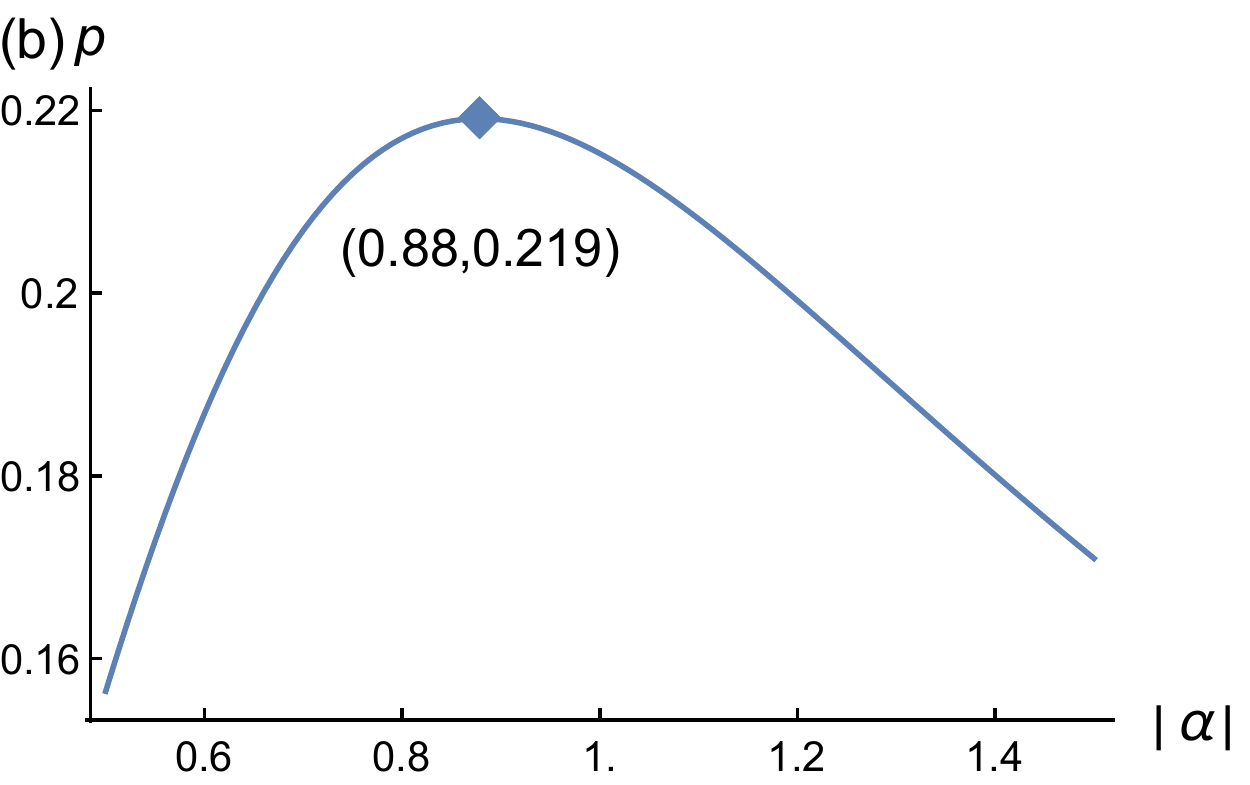}}
		\caption{Extremal regimes for mixing at a beam splitter with the coherent state $\ket{\alpha}$ in the two-site case: accepting (a) all measurement outcomes (success probability $p=1$) and (b) only $\lvert\ i-i' \rvert=\lvert\ j-j' \rvert$ (fidelity $f=1$).}
	\end{figure}

	For deterministic operation, all measurement outcomes are accepted with corresponding phase corrections to the memory. For this approach, strong coherent states are desirable for which the state transfer fidelity saturates at 0.82 (see Fig.~\subref*{fig:tradeoff_probability1}). The fidelity can be boosted by heralding on particular detection outcomes similar to the operation with ancillary $\ket{+}$ states. In particular, perfect state transfer is achievable with weak coherent states by conditioning on events where the difference in photon number between the two output ports of a beam splitter is the same between sites. Here the maximum success probability is $\sim$$0.22$ (see Fig.~\subref*{fig:tradeoff_fidelity1}). This success probability enters in the two-site protocol in the following way. The arrival time of the photon is first decoded via nonlocal parity checks with Bell pairs. For successful operation, the photonic erasure for that particular time bin must have succeeded at both sites, which happens with probability $\sim$$0.22$ for coherent state inputs.
	
	While so far we have considered a two-site array for simplicity, similar principles apply for general $N$ nodes in the network. For the deterministic operation, the state transfer fidelity of the total network state is $f(N;f_2)=(1+(N-1)(2f_2-1))/N$, where $f_2$ is the fidelity for the two-site operation. Here we have assumed that the multiple-site photonic state is a $W$ state, reflecting equal probability of the photon arriving at any node. The fidelity decreases with $N$ toward a constant $2f_2-1$. For the probabilistic operation where perfect state transfer can be obtained, the multiple-site operation will reflect the possible outcomes of $N$ probabilistic measurements.  Let $p_1$ be the success probability per site ($p_1\approx\sqrt{0.22}$). Clearly, the probability for the measurements to succeed at all $N$ nodes is $p_1^N$, which decreases exponentially with $N$. Nonetheless, even though some measurements fail, there can still be coherence between the sites with successful measurements. All coherence is lost only in the cases where all but one measurement fail or where the stellar photon is found at a site where the measurement fails. Notably, the second situation can be discriminated by the protocol by a measurement of the atomic memories. After decoding the arrival time of the photon, memories corresponding to failed photon erasure are measured to determine if a photon interacted with them. The probability of a failed total measurement over the array is
	\begin{eqnarray}
	p_{\text{fail}}&=&p_1(1-p_1)^{N-1}+\sum_{i=1}^N\binom{N}{i}(1-p_1)^ip_1^{N-i}\frac{i}{N} \nonumber \\
	&=&(1-p_1)(1+p_1(1-p_1)^{N-2})\,.
	\end{eqnarray}
	For large $N$, $p_{\text{fail}}\to(1-p_1)$. Thus, while the deterministic operation maintains coherence over the full network, probabilistic operation will in general preserve $k\leq N$ sites with success probability $\sim$$p_1$. The probability to have coherence between $k\geq2$ sites given a successful measurement is
	\begin{equation}
	p(N,k;p_1)=\binom{N}{k}p_1^k(1-p_1)^{N-k}\frac{k}{N(1-p_{\text{fail}})}\,.
	\end{equation}
	The interferometric information can be extracted from the $k$ successful sites via either the $W$ state or QFT approach described earlier.
	
	By removing the need for auxiliary receiving atoms, we circumvent atomic state detection, which may be advantageous in terms of experimental control and operation time. The requisite levels for photonic technology have already been demonstrated in tests of Bell inequality violations~\citep{Vedovato2018} and boson sampling~\citep{Spring2013}. Number-resolving detectors~\citep{Rosenberg2005} in practice do not work perfectly; in this application, only the difference in photon number between beam splitter ports is relevant, which may ease implementation. However, the simulated $X$-basis measurement is imperfect, operating with subunity fidelity or success probability. The operation could be improved by mixing more photonic states in a multiple-port beam splitter (see Appendix~\ref{sec:pPort}). Experimentally, the main limitation of the photonic state transfer approach will likely be imperfect photon detection. While a detailed study is beyond the scope of this article, we note that, in general, imperfect detection will decrease the visibility since unsuccessful events can be mistaken as successful (see Appendix~\ref{sec:imperfectDetection}).
	
	\section{Conclusions}
		
	We have analyzed refinements of our quantum network-assisted interferometry proposal~\cite{prl} for realistic broadband operation, multiple-site ($N>2$) telescope arrays, and circumvention of auxiliary atoms. The generalization to $R$ frequency bands was demonstrated, maintaining efficient scaling of entangled resources. For multiple-site operation, coherent extraction of the stellar intensity distribution by a direct implementation of the van Cittert-Zernike theorem via the quantum Fourier transform was shown to yield significant improvement in signal-to-noise ratio compared to visibility readout and classical postprocessing. This result was obtained assuming perfect operation. An interesting extension of this work would be to study the effect of a noisy QFT on the imaging capabilities. We also analyzed an implementation of our proposal using direct photon-memory interaction. In particular, a photonic $X$-basis measurement is accomplished by mixing with ancillary photonic states at beam splitters followed by photon counting. The scheme then minimizes the need for atomic measurement, at the expense of introducing photon-resolving detectors, beam splitters, and ancillary photonic states. The considerations in this article reinforce the power of quantum networks as a platform for astronomical interferometry and provide a path toward implementation.
	
	\begin{acknowledgments}
	We thank A. Aspuru-Guzik, M. Bhaskar, F. Brand\~ao, M. Christandl, I. Chuang, I. Cirac, J. Stark, C. Stubbs, H. Zhou, and P. Zoller for illuminating discussions and useful comments. This work was supported by the National Science Foundation (NSF), the Center for Ultracold Atoms, an NSF Graduate Research Fellowship (E.T.K.), a Vannevar Bush Faculty Fellowship, ERC Grant Agreement No. 337603, the Danish Council for Independent Research (Sapere Aude), Qubiz -- Quantum Innovation Center, and VILLUM FONDEN via the QMATH Centre of Excellence (Grant No. 10059).
	\end{acknowledgments}
	
	\appendix
	
	\section{Multiple-port beam splitter} \label{sec:pPort}
	
	The photon-detection-based state transfer can be viewed like a quantum teleportation~\citep{Kok2010}. The entangled resource state is the stellar photon entangled with the quantum memory, and the beam splitter with the ancillary input state realizes an (imperfect) Bell measurement. Upon applying a corrective unitary based on the measurement outcomes registered by the detectors, the state of the stellar light is transferred to memory. The operation can be improved with a better Bell measurement, but a no-go theorem~\citep{Lutkenhaus1999} excludes a perfect deterministic measurement for single-photon qubits and linear optics. Nonetheless, an approximation arbitrarily close to ideal may be constructed, like in the famous protocol of \citet{Knill2001}. The idea is to better hide the which-path information of the photon by sending it through a multiple-port beam splitter, which may be constructed from layers of ordinary two-port beam splitters. A $P+1$ port realizes the QFT unitary $U_{\text{QFT}}(P+1)$ (Eq.~\eqref{eq:QFT}), which generalizes the Hadamard gate and provides a change of basis. Generalizing the above procedure, the photon reflected off the memory cavities is input along with $P$ ancillary photonic states into a $(P+1)$-port beam splitter. Again, the detector measurement outcomes determine the phase correction to apply to complete state transfer.
	
	\section{Imperfect photon counting} \label{sec:imperfectDetection}
	
\begin{figure}
		\centering
		\includegraphics[width=0.5\textwidth]{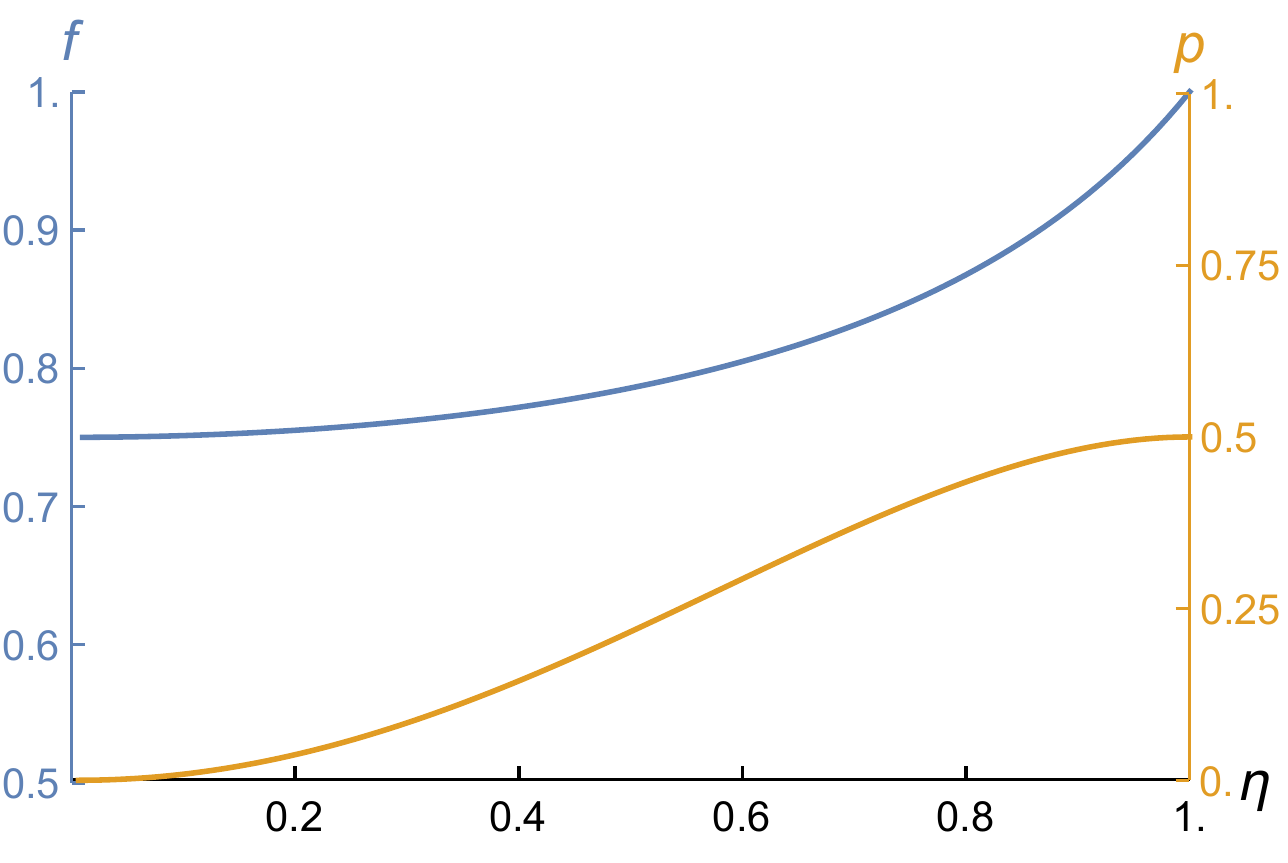}
		\caption{Fidelity $f$ and success probability $p$ of photonic state transfer when mixing with an ancillary $\ket{+}$ state and detecting photons with efficiency $\eta$. Measurement outcomes of zero or two total photons are postselected out. The performance of the operation drops with $\eta$.
			\label{fig:lossy}}
	\end{figure}

	The phase correction necessary to complete photon-detection-based state transfer depends on the measurement outcome. Detector errors decrease the coherence of the target qubit through the application of incorrect recovery operations. Ultimately, the visibility is reduced. We consider this effect in the case of operation with ancillary $\ket{+}$ states. The inefficient detectors are modeled as beam splitters with transmission amplitude $\eta$. The signal comes in one port and vacuum in the other, followed by perfect detection of one output mode while the lossy mode is traced over. As expected, the fidelity and success probability of the state transfer decrease with $\eta$ (see Fig.~\ref{fig:lossy}).
	
	\clearpage
	\bibliography{pra}

\end{document}